\documentclass{article}
\usepackage{amsmath, amssymb, geometry, hyperref}
\newgeometry{left = 3cm, right = 3cm}
\usepackage{hyperref}
\hypersetup{colorlinks=true,linkcolor=black,urlcolor=blue,citecolor=black}
\usepackage[round]{natbib}
\newcommand{\st}[1]{\texttt{#1}}
\author{Diego Ciccia\thanks{Northwestern University, Kellogg School of Management. diego.ciccia@kellogg.northwestern.edu.}}\date{}
\title{A Short Note on Event-Study Synthetic Difference-in-Differences Estimators}

\begin{document}

\maketitle

\abstract{I propose an event study extension of Synthetic Difference-in-Differences (SDID) estimators. I show that, in simple and staggered adoption designs, estimators from \cite{arkhangelsky2021synthetic} can be disaggregated into dynamic treatment effect estimators, comparing the lagged outcome differentials of treated and synthetic controls to their pre-treatment average. Estimators presented in this note can be computed using the \st{sdid\_event} Stata package\footnote{The package is hosted on \href{https://github.com/DiegoCiccia/sdid/tree/main/sdid_event}{Github}.}.}
\paragraph{Adapting SDID to event study analysis.}

In what follows, I use the notation from \cite{clarke2023synthetic} to present the estimation procedure for event-study Synthetic Difference-in-Differences (SDID) estimators. In a setting with $N$ units observed over $T$ periods, $N_{tr} < N$ units receive treatment $D$ starting from period $a$, where $1 < a \leq T$. The treatment $D$ is binary, i.e. $D \in \lbrace 0,1\rbrace$, and it affects some outcome of interest $Y$. The outcome and the treatment are observed for all $(i,t)$ cells, meaning that the data has a balanced panel structure. 

Henceforth, values of $a$ are referred to as \emph{cohorts} or \emph{adoption periods}. The values of $a$ are collected in $A$, i.e. the adoption date vector. For the sake of generality, we assume that $|A| > 1$, meaning that groups start receiving the treatment at different periods. The case with no differential timing can be simply retrieved by considering one cohort at a time. Time periods are indexed by $t \in \lbrace 1, ..., T\rbrace$, while units are indexed by $i \in \lbrace 1, ... N\rbrace$. Without loss of generality, the first $N_{co} = N - N_{tr}$ units are the never-treated group. As for the treated units, let $I^a$ be the subset of $\lbrace N_{co} +1, ..., N\rbrace$ containing the indices of units in cohort $a$.  Lastly, we denote with $N^{a}_{tr}$ and $T^{a}_{tr}$ the number of units in cohort $a$ and the number of periods from the the onset of the treatment in the same cohort to end of the panel, respectively. These two cohort-specific quantities can be aggregated into $T_{post}$ from \cite{clarke2023synthetic}, i.e. the total number of post treatment periods of all the units in every cohort. Namely,

\begin{equation}
T_{post} = \sum_{a \in A} N^{a}_{tr}T^{a}_{tr}
\end{equation}

is the sum of the products of $N^{a}_{tr}$ and $T^{a}_{tr}$ across all $a \in A$.

\paragraph{Disaggregating $\hat{\tau}^{sdid}_a$.}

The cohort-specific SDID estimator from \cite{arkhangelsky2021synthetic} can be rearranged as follows:

\begin{equation}
\hat{\tau}^{sdid}_a = \frac{1}{T^a_{tr}} \sum_{t = a}^T \left( \frac{1}{N^{a}_{tr}} \sum_{i \in I^a} Y_{i,t} - \sum_{i = 1}^{N_{co}} \omega_i Y_{i,t}\right) -  \sum_{t = 1}^{a-1} \left( \frac{1}{N^{a}_{tr}} \sum_{i \in I^a} \lambda_t Y_{i,t} - \sum_{i = 1}^{N_{co}}\omega_i \lambda_t  Y_{i,t}\right)
\end{equation}

where $\lambda_t$ and $\omega_i$ are the optimal weights chosen to best approximate the pre-treatment outcome evolution of treated and (synthetic) control units. $\tau^{sdid}_a$ compares the average outcome difference of treated in cohort $a$ and never-treated before and after the onset of the treatment. In doing so, $\tau^{sdid}_a$ encompasses all the post-treatment periods. As a result, it is possible to estimate the treatment effect $\ell$ periods after the adoption of the treatment, with $\ell \in \lbrace 1,..., T^a_{post} \rbrace$, via a simple disaggregation of $\tau^{sdid}_a$ into the following event-study estimators:

\begin{equation}
\hat{\tau}^{sdid}_{a, \ell} = \frac{1}{N^{a}_{tr}} \sum_{i \in I^a} Y_{i,a-1+\ell} - \sum_{i = 1}^{N_{co}} \omega_i Y_{i,a-1+\ell} -  \sum_{t = 1}^{a-1} \left( \frac{1}{N^{a}_{tr}} \sum_{i \in I^a} \lambda_t Y_{i,t} - \sum_{i = 1}^{N_{co}}\omega_i \lambda_t  Y_{i,t}\right)
\end{equation}

This estimator is very similar to those proposed by \cite{borusyak2024revisiting}, \cite{liu2024practical} and \cite{gardner2022two}, when the design is a canonical DiD \citep{de2023difference}. The only difference lies in the fact that the outcomes are weighted via unit-time specific weights. Notice that by construction

\begin{equation}\label{dis1}
\hat{\tau}^{sdid}_a = \frac{1}{T^a_{tr}} \sum_{\ell = 1}^{T^a_{tr}} \hat{\tau}^{sdid}_{a, \ell}
\end{equation}

that is, $\hat{\tau}^{sdid}_a$ is the sample average of the cohort-specific dynamic estimators $\hat{\tau}^{sdid}_{a, \ell}$.

\paragraph{Aggregating $\hat{\tau}^{sdid}_{a,\ell}$ estimators into event-study estimates.}

Let $A_{\ell}$ be the subset of cohorts in $A$ such that  $a - 1 + \ell \leq T$, i.e. such that their $\ell$-th dynamic effect can be computed, and let 

\begin{equation}
N^{\ell}_{tr} = \sum_{a \in A_\ell} N^a_{tr}
\end{equation}

denote the number of units in cohorts where the $\ell$-th dynamic effect can be estimated. We can use this notation to aggregate the cohort-specific dynamic effects into a single estimator. Let

\begin{equation}
\hat{\tau}^{sdid}_\ell = \sum_{a \in A_{\ell}} \frac{N^a_{tr}}{N^{\ell}_{tr}} \hat{\tau}^{sdid}_{a,\ell}
\end{equation}

denote the weighted sum of the cohort-specific treatment effects $\ell$ periods after the onset of the treatment, with weights corresponding to the relative number of groups participating into each cohort. This estimator aggregates the cohort-specific treatment effects, upweighting more representative cohorts in terms of units included. 

As in Equation \ref{dis1}, $\hat{\tau}^{sdid}_\ell$ can also be retrieved via disaggregation of another estimator from \cite{clarke2023synthetic}. Let $T_{tr} = \max_{a \in A} T^{a}_{tr}$ be the maximum number of post-treatment periods across all cohorts. Equivalently, $T_{tr}$ can also be defined as the number of post-treatment periods of the earliest treated cohort. Then, one can show that
\begin{equation}\label{dis2}
\widehat{ATT} = \dfrac{1}{T_{post}} \sum_{\ell = 1}^{T_{tr}} N^{\ell}_{tr} \hat{\tau}^{sdid}_\ell
\end{equation}
that is, the $\widehat{ATT}$ estimator from \cite{clarke2023synthetic} is a weighted average of the event study estimators $\hat{\tau}^{sdid}_\ell$, with weights proportional to the number of units for which the $\ell$-th effect can be computed. 

\paragraph{Mapping with estimates from sdid\_event.} Estimators presented in this note can be computed using the \st{sdid\_event} Stata package. The baseline output table of \st{sdid\_event} reports the estimates of the ATT from \st{sdid} and $\hat{\tau}^{sdid}_\ell$ for $\ell \in \lbrace 1, ..., T_{tr} \rbrace$. If the command is run with the \st{disag} option, the output also includes a table with the cohort-specific treatment effects. Namely, the program returns the estimates of  $\hat{\tau}^{sdid}_a$ and $\hat{\tau}^{sdid}_{a,\ell}$, whereas the former can also be retrieved from the \st{e(tau)} matrix in \st{sdid}.

\subparagraph{Proof of Equation \ref{dis2}.} 
Let $T^{a}_{post}$ denote the total number of post-treatment periods across all units in cohort $a$.

\[
\begin{array}{rcl}
\widehat{ATT} &=&\sum_{a \in A} \dfrac{T^{a}_{post}}{T_{post}}  \hat{\tau}^{sdid}_a \\
&=& \dfrac{1}{T_{post}} \sum_{a \in A} N^{a}_{tr} T^{a}_{tr}  \hat{\tau}^{sdid}_a \\
&=& \dfrac{1}{T_{post}} \sum_{a \in A} \sum_{\ell = 1}^{T^a_{tr}} N^{a}_{tr} \hat{\tau}^{sdid}_{a, \ell} \\
&=& \dfrac{1}{T_{post}} \sum_{\ell = 1}^{T_{tr}} \sum_{a \in A_{\ell}} N^{a}_{tr} \hat{\tau}^{sdid}_{a, \ell} \\
&=& \dfrac{1}{T_{post}} \sum_{\ell = 1}^{T_{tr}} N^{\ell}_{tr} \hat{\tau}^{sdid}_\ell
\end{array}
\]
where the first equality comes from the definition of $\widehat{ATT}$ in \cite{clarke2023synthetic}, the second equality from the definition of $T^{a}_{post}$, the third equality from Equation \ref{dis1}, the fourth equality from the fact that the sets $\lbrace (a, \ell): a \in A, 1 \leq \ell \leq T^{a}_{tr} \rbrace$ and $\lbrace (a, \ell): 1 \leq \ell \leq T_{tr}, a \in A_{\ell}\rbrace$ are equal and the fifth equality from the definition of $\hat{\tau}^{sdid}_\ell$.

\bibliographystyle{plainnat}
\bibliography{sdid_event}

\end{document}